\documentclass[12pt,epsfig]{article}
\usepackage{amssymb,epsfig}

\makeatletter
\@addtoreset{equation}{section}

\newcommand{\cQ}{{\cal Q}}

\newcommand{\nn}{\nonumber \\}

\def\be{\begin{equation}}
\def\ee{\end{equation}}
\def\beq{\begin{equation}}
\def\eeq{\end{equation}}
\def\bea{\begin{eqnarray}}
\def\eea{\end{eqnarray}}
\begin{document}

\begin{titlepage}
\hfill
\vbox{
    \halign{#\hfil         \cr
           TAUP-2796-05\cr
           hep-th/0502187  \cr
           } 
      }  
\vspace*{20mm}

\begin{center}
{\Large {\bf On Type II 
    Strings in Two Dimensions}\\} 
\vspace*{15mm}

{\sc Harald Ita}
\footnote{e-mail: {\tt h.ita@swan.ac.uk}},
{\sc Harald Nieder}
\footnote{e-mail: {\tt harald@post.tau.ac.il}}, 
{\sc Yaron Oz} 
\footnote{e-mail: {\tt yaronoz@post.tau.ac.il}}\\

\vspace*{1cm} 
{\it ${}^1$ Department of Physics\\
University of Wales Swansea\\
Swansea, SA 8PP, UK\\}

\vspace*{1cm} 
{\it ${}^{2,3}$ Raymond and Beverly Sackler Faculty of Exact Sciences\\
School of Physics and Astronomy\\
Tel-Aviv University , Ramat-Aviv 69978, Israel\\}

\end{center}

\vspace*{8mm}
 
\begin{abstract}

We consider type IIA/B strings in two-dimensions and
their projection with respect to the nilpotent space-time
supercharge.
Based on the ground ring structure,
we propose a duality between
perturbed type II strings and
the topological B-model on deformed Calabi-Yau
singularities.
Depending on the type II spectra, one has either the conifold
or the   suspended pinch point geometry.
Using the corresponding quiver gauge theory, obtained by D-branes
wrapping in the resolved suspended pinch point geometry, we 
propose the all orders perturbative partition function.

\end{abstract}
\vskip 1cm

February  2005

\end{titlepage}

\setcounter{footnote}{0}

\section{Introduction}

There are by now several examples of equivalence relations 
between non-critical string theories and closed topological  
B-model on non-compact Calabi-Yau 3-folds \cite{Aganagic:2003qj}.
The basic ingredient in the relation is the string theory ground ring,
whose (complexified) defining equations are identified
with the Calabi-Yau space.
Typically, the  ground ring  defines a Calabi-Yau singularity, and its
deformation corresponds to turning on a string perturbation, such as 
a cosmological constant or RR flux.

The
 non-critical $c=1$ bosonic string at the self-dual radius,
has been related to 
the topological B-model on the deformed conifold \cite{Ghoshal:1995wm,Dijkgraaf:2003xk}.
Type 0 non-critical $\hat{c}=1$ strings have 
been related to 
the topological B-model on certain $Z_2$ 
quotients of the conifold \cite{Ita:2004yn} (for further
related work see \cite{Danielsson:2004ti,Park:2004yc,Hyun:2005fq}).
The partition functions of the non-critical
strings 
have been matched perturbatively 
with those of the topological B-models.

By wrapping D-branes on the resolved   Calabi-Yau singularities, one obtains
a UV description of four-dimensional ${\cal N}=1$ quiver gauge theories.
Their IR physics is obtained by running duality cascades on the field
theory side, or by the geometric transition in the topological
picture.
The partition function of the non-critical strings is
identified with the glueball F-terms of the
gauge theory \cite{Dijkgraaf:2003xk}.

In this paper we will propose
a duality between
type II strings in two-dimensions and
topological B-model on deformed Calabi-Yau singularities.
We will find two types of singularities depending on the choice of the type II spectra.
One will be the conifold and the other will be the suspended pinch point
singularity. 
These Calabi-Yau singularities describe the complexified ground ring relations. 
On the type II string side the deformations of the singularities
correspond to turning on RR field background.
Using the corresponding quiver gauge theory, obtained by D-branes
wrapping in the resolved geometry, we 
propose the all orders perturbative partition function, which coincides
with that  of the non-critical $c=1$ bosonic string.

The paper is organized as follows.
In section 2 we will study the two-dimensional type IIA and  type IIB strings.
We will analyze their spectra, ground ring structure
and their deformations, and the one-loop partition sums.
We will also consider projections of these theories by the space-time
supercharge $\cQ$, which is a nilpotent
BRST like operators. 
We will propose a description of perturbed  type II strings
as a topological B-model, on the (deformed) conifold or
suspended pinch point singularities.
In section 3 we will discuss the four-dimensional
${\cal N}=1$ quiver gauge theory obtained
from D-branes wrapping in the  suspended pinch point geometry. Using its IR
physics, we will suggest that the partition function of the perturbed
type II strings and the corresponding topological
B-model, are given by the partition function of
the $c=1$ bosonic string at the self-dual radius.

\section{$2d$ type II Strings}

The two-dimensional fermionic string is described in the superconformal
gauge by a matter superfield $X$ and the super
Liouville field $\Phi$. In components they take the form
\bea
X &=& x+\theta\psi_x+\bar\theta\bar\psi_x+\theta\bar\theta F_x \ ,
 \nonumber\\
\Phi &=&\phi+\theta\psi_l+\bar\theta\bar\psi_l+\theta\bar\theta F_l \ .
\eea
The field $X$ is free, while $\Phi$ has a background charge $Q=2$.
The relevant OPEs are
\beq
\phi(z)\phi(0)=x(z)x(0)\sim-log(z),\quad \psi_{l,x}(z)
\psi_{l,x}(0)\sim 1/z \ .
\eeq

Let us consider the GSO projections
of the type IIA/B theories and the respective physical spectra.
We will consider $x$ compactified on a circle of radius $R=2$, and $\alpha'=2$, which
is the point consistent with ${\cal N}=2$ supersymmetry on the worldsheet \cite{Kutasov:1990ua}.
The GSO projection is imposed 
by requiring the physical vertex operators to be 
local with respect to the space-time supercharges. For the
type IIB theory the current will be given by
\begin{equation}
\label{Sp}
S^+(z)=e^{-\frac{\sigma}{2}+\frac{iH}{2}+iX_L} \ ,
\end{equation}
in the left moving sector, and by
\begin{equation}
\label{bSp}
\bar{S}^+(\bar{z})=e^{-\frac{\bar{\sigma}}{2}+\frac{i\bar{H}}{2}+iX_R} \ ,
\end{equation}
in the right moving sector.
For the type IIA theory, we require locality with respect to
$S^+(z)$ defined by (\ref{Sp}) 
in the left moving sector, and with respect to
\begin{equation}
\label{bSm}
\bar{S}^-(\bar{z})=e^{-\frac{\bar{\sigma}}{2}-\frac{i\bar{H}}{2}-iX_R} \ ,
\end{equation}
in the right moving sector.

$\sigma$ is the bosonized superconformal ghost, $\beta\gamma = \partial \sigma$, and
$H$ is obtained by bosonizing the fermions $\psi_x\psi_l=\partial H$.
In our conventions the conformal dimensions are as follows
\begin{eqnarray}
\left[e^{ikX}\right]=\frac{k^2}{2},\quad
\left[e^{-l\sigma}\right]=-\frac{l^{2}}{2}+l,\quad \left[e^{inH}\right]=\frac{n^{2}}{2},
\quad
\left[e^{\alpha \phi}\right]=-\frac{1}{2}\alpha(\alpha-2) \ .
\end{eqnarray}

\subsection{Type IIB}
\label{secIIb}

The spectrum is determined by requiring locality with respect
to the supercharges, level matching and mutual locality. However,
in the case at hand this does not define the spectrum
uniquely. In addition we have the freedom to choose the lowest lying
NSNS state to be a momentum or a winding state. The theory
perturbed by these lowest lying states is then either type II on the trumpet geometry,
for the momentum perturbation, or type II on the cigar geometry, for the
winding perturbation. The latter case was analyzed 
in \cite{Murthy:2003es}. For the sake of completeness this section
will cover all the available spectra.

\noindent
{\it ``Momentum Background''}

\noindent

Due to the absence of transverse excitations the spectrum is very simple.
In the NS sector the tachyon vertex operators are 
\begin{equation}
\label{Tk}
T_k = e^{-\sigma +ikX+\beta \phi} \ . 
\end{equation}
Locality with respect to the space-time supercharge projects onto
half-integer values for the momentum in the $x$-direction in 
both the left moving and the right moving sector. Thus,
\beq
X: k_L, k_R \in Z +\frac{1}{2} \ .
\eeq
In order to obtain the closed 
string states we need to combine the left and right moving states in such
a way that the momentum along the Liouville direction, which is non-compact,
is the same in both sectors. 

The half-integer moding does not allow for
winding states in the NSNS sector. This is because 
locality with respect to the lowest lying NSNS momentum state implies that
the left and 
right moving momenta are related by \cite{Kutasov:1991pv}
\begin{equation}
k_L-k_R \in 2Z \ ,
\end{equation}
and we can write $k_{L,R}=m \pm n +\frac{1}{2}$.
Level matching implies that $k_L^2=k_R^2$, and the only possible solution
is $n=0$.

In the Ramond  sector we have the chiral vertex operators 
\begin{equation}
\label{Vk}
V_k=e^{-\frac{\sigma}{2} +\frac{i}{2} \epsilon H+ ikX +\beta \phi} \ ,
\end{equation}   
and the GSO projection implies $k \in Z $ for $\epsilon=1$ and
$k \in Z + \frac{1}{2}$ for $\epsilon=-1$, in both left and right sectors.

Before imposing the Dirac equation constraint we find the spectrum: 
\begin{eqnarray}
\label{iib}
NSNS &  e^{-\sigma-\bar{\sigma}+i\left(n+\frac{1}{2}\right)\left(
X_L+X_R\right)} \ , & n \in Z \ , \\ \nonumber
RR,\, \epsilon_L=\epsilon_R=1 &
e^{-\frac{\sigma}{2}-\frac{\bar{\sigma}}{2}+\frac{i}{2}\left(H+\bar{H}\right)
+im\left(X_L+X_R\right)} \ , & m \in Z \ , \\ \nonumber
& e^{-\frac{\sigma}{2}-\frac{\bar{\sigma}}{2}+\frac{i}{2}\left(H+\bar{H}\right)
+im\left(X_L-X_R\right)} \ , & m \ne 0 \ , \\ \nonumber
RR,\, \epsilon_L=\epsilon_R=-1 &
e^{-\frac{\sigma}{2}-\frac{\bar{\sigma}}{2}-\frac{i}{2}\left(H+\bar{H}\right)
+i\left(n+\frac{1}{2}\right) \left(X_L+X_R\right)} \ , & n \in Z \ , \\ \nonumber
NSR & e^{-\sigma-\frac{\bar{\sigma}}{2}-i\frac{\bar{H}}{2}
+i\left(n + \frac{1}{2} \right) \left(X_L-X_R\right)} \ , & n \in Z \ , \\ \nonumber
& e^{-\frac{\sigma}{2}-\bar{\sigma}-i\frac{H}{2}
+i\left(n + \frac{1}{2} \right) \left(X_L-X_R\right)} \ , & n \in Z \ ,
\end{eqnarray}
where we have omitted the Liouville dressing.
The Liouville dressing is determined by
requiring conformal invariance of the integrated vertex operators. In our
conventions this requires the coefficient $\beta$ to be a solution to the equation
\begin{equation}
\frac{k^2}{2}-\frac{1}{2}\beta\left(\beta-2\right)=\frac{1}{2} \ .
\end{equation}
Furthermore the locality constraint found in \cite{Seiberg:1990eb} requires
$\beta \le \frac{Q}{2}$.
Note, however, that here BRST invariance does not follow from
conformal invariance alone \cite{DiFrancesco:1991ud}.
Invariance with respect to the susy BRST operator $Q_{susy} = \oint \gamma T_F$ imposes
an additional constraint, equivalent to the space-time Dirac equation.  
Thus, the physical states satisfy in addition
$\beta -\frac{Q}{2}=-\epsilon k$.
Combining this with the relation following from
conformal invariance we conclude that for physical momenta 
in the Ramond sector $\left|k \right|=\epsilon k$.

The spectrum consistent with all the above constraints is given by
\begin{eqnarray}
\label{iibd}
NSNS &  e^{-\sigma-\bar{\sigma}+i\left(n+\frac{1}{2}\right)\left(
X_L+X_R\right)} \ , & n \in Z \ , \\ \nonumber
RR,\, \epsilon_L=\epsilon_R=1 &
e^{-\frac{\sigma}{2}-\frac{\bar{\sigma}}{2}+\frac{i}{2}\left(H+\bar{H}\right)
+im\left(X_L+X_R\right)} \ , & m \ge 0 \ , \\ \nonumber
RR,\, \epsilon_L=\epsilon_R=-1 &
e^{-\frac{\sigma}{2}-\frac{\bar{\sigma}}{2}-\frac{i}{2}\left(H+\bar{H}\right)
+i\left(n+\frac{1}{2}\right) \left(X_L+X_R\right)} \ , & n < 0 \ , \\ \nonumber
NSR & e^{-\sigma-\frac{\bar{\sigma}}{2}-i\frac{\bar{H}}{2}
+i\left(n + \frac{1}{2} \right) \left(X_L-X_R\right)} \ , & n \ge 0 \ , \\ \nonumber
& e^{-\frac{\sigma}{2}-\bar{\sigma}-i\frac{H}{2}
+i\left(n + \frac{1}{2} \right) \left(X_L-X_R\right)} \ , & n < 0 \ .
\end{eqnarray}

\noindent
{\it ``Winding Background''}

\noindent
If we choose instead to have winding modes in the NSNS sector we
find the following spectrum \cite{Murthy:2003es,Seiberg:2005bx}
 \begin{eqnarray}
\label{iibw}
NSNS &  e^{-\sigma-\bar{\sigma}+i\left(n+\frac{1}{2}\right)\left(
X_L-X_R\right)} \ , & n \in Z \ , \\ \nonumber
RR,\, \epsilon_L=\epsilon_R=1 &
e^{-\frac{\sigma}{2}-\frac{\bar{\sigma}}{2}+\frac{i}{2}\left(H+\bar{H}\right)
+im\left(X_L+X_R\right)} \ , & m \in Z \ , \\ \nonumber
& e^{-\frac{\sigma}{2}-\frac{\bar{\sigma}}{2}+\frac{i}{2}\left(H+\bar{H}\right)
+im\left(X_L-X_R\right)} \ , & m \ne 0 \ , \\ \nonumber
RR,\, \epsilon_L=\epsilon_R=-1 &
e^{-\frac{\sigma}{2}-\frac{\bar{\sigma}}{2}-\frac{i}{2}\left(H+\bar{H}\right)
+i\left(n+\frac{1}{2}\right) \left(X_L-X_R\right)} \ , & n \in Z \ , \\ \nonumber
NSR & e^{-\sigma-\frac{\bar{\sigma}}{2}-i\frac{\bar{H}}{2}
+i\left(n + \frac{1}{2} \right) \left(X_L+X_R\right)} \ , & n \in Z \ , \\ \nonumber
& e^{-\frac{\sigma}{2}-\bar{\sigma}-i\frac{H}{2}
+i\left(n + \frac{1}{2} \right) \left(X_L+X_R\right)} \ , & n \in Z \ .
\end{eqnarray}
Again, after imposing the Dirac condition the physical spectrum is given by
\begin{eqnarray}
\label{iibwd}
NSNS &  e^{-\sigma-\bar{\sigma}+i\left(n+\frac{1}{2}\right)\left(
X_L-X_R\right)} \ , & n \in Z \ , \\ \nonumber
RR,\, \epsilon_L=\epsilon_R=1 &
e^{-\frac{\sigma}{2}-\frac{\bar{\sigma}}{2}+\frac{i}{2}\left(H+\bar{H}\right)
+im\left(X_L+X_R\right)} \ , & m \ge 0 \ , \\ \nonumber
NSR & e^{-\sigma-\frac{\bar{\sigma}}{2}-i\frac{\bar{H}}{2}
-i\left(n + \frac{1}{2} \right) \left(X_L+X_R\right)} \ , & n \ge 0 \ , \\ \nonumber
& e^{-\frac{\sigma}{2}-\bar{\sigma}-i\frac{H}{2}
-i\left(n + \frac{1}{2} \right) \left(X_L+X_R\right)} \ , & n \ge 0 \ .
\end{eqnarray}

\subsection{Type IIA} 
\label{secIIa}

The type IIA NSNS sector is the same as the type IIB theory one.
In the right moving Ramond sector we find a modification:
operators with $\epsilon_R=1$ have momentum $k_R \in Z+\frac{1}{2}$, whereas
operators with $\epsilon_R=-1$ have momentum $k_R \in Z$.

\noindent
{\it ``Momentum Background''}

\noindent
The spectrum is given by
\begin{eqnarray}
\label{iiam}
NSNS &  e^{-\sigma-\bar{\sigma}+i\left(n+\frac{1}{2}\right)\left(
X_L+X_R\right)} \ , & n \in Z \ , \\ \nonumber
RR,\, \epsilon_L=1, \epsilon_R=-1 &
e^{-\frac{\sigma}{2}-\frac{\bar{\sigma}}{2}+\frac{i}{2}\left(H-\bar{H}\right)
+im\left(X_L-X_R\right)} \ , & m \in Z \ , \\ \nonumber
& e^{-\frac{\sigma}{2}-\frac{\bar{\sigma}}{2}+\frac{i}{2}\left(H-\bar{H}\right)
+im\left(X_L+X_R\right)} \ , & m \ne 0 \ , \\ \nonumber
RR,\, \epsilon_L=-1, \epsilon_R=1 &
e^{-\frac{\sigma}{2}-\frac{\bar{\sigma}}{2}-\frac{i}{2}\left(H-\bar{H}\right)
+i\left(n+\frac{1}{2}\right) \left(X_L+X_R\right)} \ , & n \in Z \ , \\ \nonumber
NSR & e^{-\sigma-\frac{\bar{\sigma}}{2}+i\frac{\bar{H}}{2}
+i\left(n + \frac{1}{2} \right) \left(X_L-X_R\right)} \ , & n \in Z \ , \\ \nonumber
& e^{-\frac{\sigma}{2}-\bar{\sigma}-i\frac{H}{2}
+i\left(n + \frac{1}{2} \right) \left(X_L-X_R\right)} \ , & n \in Z \ .
\end{eqnarray}
After imposing the Dirac equation we are left with 
\begin{eqnarray}
\label{iiamd}
NSNS &  e^{-\sigma-\bar{\sigma}+i\left(n+\frac{1}{2}\right)\left(
X_L+X_R\right)} \ , & n \in Z \ , \\ \nonumber
RR,\, \epsilon_L=1, \epsilon_R=-1 &
e^{-\frac{\sigma}{2}-\frac{\bar{\sigma}}{2}+\frac{i}{2}\left(H-\bar{H}\right)
+im\left(X_L-X_R\right)} \ , & m \ge 0 \ , \\ \nonumber
NSR & e^{-\sigma-\frac{\bar{\sigma}}{2}+i\frac{\bar{H}}{2}
+i\left(n + \frac{1}{2} \right) \left(X_L-X_R\right)} \ , & n < 0 \ , \\ \nonumber
& e^{-\frac{\sigma}{2}-\bar{\sigma}-i\frac{H}{2}
+i\left(n + \frac{1}{2} \right) \left(X_L-X_R\right)} \ , & n < 0 \ .
\end{eqnarray}

\noindent
{\it ``Winding Background''}

\noindent
As in the type IIB case we can also start with winding modes in the NSNS
sector. The consistent spectrum is then easily derived to be given by
\begin{eqnarray}
\label{iiaw}
NSNS &  e^{-\sigma-\bar{\sigma}+i\left(n+\frac{1}{2}\right)\left(
X_L-X_R\right)} \ , & n \in Z \ , \\ \nonumber
RR,\, \epsilon_L=1, \epsilon_R=-1 &
e^{-\frac{\sigma}{2}-\frac{\bar{\sigma}}{2}+\frac{i}{2}\left(H-\bar{H}\right)
+im\left(X_L-X_R\right)} \ , & m \in Z \ , \\ \nonumber
& e^{-\frac{\sigma}{2}-\frac{\bar{\sigma}}{2}+\frac{i}{2}\left(H-\bar{H}\right)
+im\left(X_L+X_R\right)} \ , & m \ne 0 \ , \\ \nonumber
RR,\, \epsilon_L=-1, \epsilon_R=1 &
e^{-\frac{\sigma}{2}-\frac{\bar{\sigma}}{2}-\frac{i}{2}\left(H-\bar{H}\right)
+i\left(n+\frac{1}{2}\right) \left(X_L-X_R\right)} \ , & n \in Z \ , \\ \nonumber
NSR & e^{-\sigma-\frac{\bar{\sigma}}{2}+i\frac{\bar{H}}{2}
+i\left(n + \frac{1}{2} \right) \left(X_L+X_R\right)} \ , & n \in Z \ , \\ \nonumber
& e^{-\frac{\sigma}{2}-\bar{\sigma}-i\frac{H}{2}
+i\left(n + \frac{1}{2} \right) \left(X_L+X_R\right)} \ , & n \in Z \ .
\end{eqnarray}
Similarly, the Dirac condition now leads to the following spectrum
\begin{eqnarray}
\label{iiawd}
NSNS &  e^{-\sigma-\bar{\sigma}+i\left(n+\frac{1}{2}\right)\left(
X_L-X_R\right)} \ , & n \in Z \ , \\ \nonumber
RR,\, \epsilon_L=1, \epsilon_R=-1 &
e^{-\frac{\sigma}{2}-\frac{\bar{\sigma}}{2}+\frac{i}{2}\left(H-\bar{H}\right)
+im\left(X_L-X_R\right)} \ , & m \ge 0 \ , \\ \nonumber
RR,\, \epsilon_L=-1, \epsilon_R=1 &
e^{-\frac{\sigma}{2}-\frac{\bar{\sigma}}{2}-\frac{i}{2}\left(H-\bar{H}\right)
-i\left(n+\frac{1}{2}\right)\left(X_L-X_R\right)} \ , & n \ge 0 \ , \\ \nonumber
NSR & e^{-\sigma-\frac{\bar{\sigma}}{2}+i\frac{\bar{H}}{2}
+i\left(n + \frac{1}{2} \right) \left(X_L+X_R\right)} \ , & n \ge 0 \ , \\ \nonumber
& e^{-\frac{\sigma}{2}-\bar{\sigma}-i\frac{H}{2}
-i\left(n + \frac{1}{2} \right) \left(X_L+X_R\right)} \ , & n \ge 0 \ .
\end{eqnarray}
We note in passing that different conventions for the supercharges
which replace $S^+ \leftrightarrow S^-$ would result in a set of equivalent
theories.

\subsection{Ground Ring}

The
spin zero ghost number zero BRST invariant operators generate a commutative,
 associative
ring
\beq
\label{BRS}
{\cal O}(z){\cal O}'(0)\sim {\cal O}''(0)+\{Q_B,\dots\} \ ,
\eeq
called the ground ring, where $Q_B$ is the $N=1$ BRST operator.

In the following we will analyze the ground ring for the type II.
The relevant BRST analysis 
has been performed in  \cite{Itoh:1991ix,Bouwknegt:1991va,Bouwknegt:1991am}.

Consider the left sector.
The chiral BRST cohomology of dimension zero and ghost number zero is given by the 
infinite set of states
$\Psi_{(r,s)}$ with $r,s$ negative integers
\beq
\Psi_{(r,s)}\sim O_{r,s} e^{\left(ik_{r,s} x_L-p_{r,s}\Phi_L\right)} \ .
\eeq
The Liouville and matter momentum are given by 
\beq
k_{r,s}=\frac{1}{2}(r-s),\quad p_{r,s}=\frac{1}{2}(r+s+2) \ .
\eeq
The operators $\Psi_{(r,s)}$  are in the NS-sector if 
$k_{r,s}=(r-s)/2$ takes integer values, and in the R-sector if it takes half integer values. 

Of particular relevance for us are the R-sector operators:
\begin{eqnarray}\label{chiralring}
x(z)&\!\equiv\!&\Psi_{(-1,-2)}(z) 
\!=\!
\left( e^{-{i\over 2}H}e^{-{1\over 2}\sigma}
-{1\over\sqrt{2}}e^{{i\over 2}H}\partial\xi e^{-{3\over 2}\sigma}
\right) e^{{i\over 2}x-{1\over 2}\phi} \ , \nn
y(z)&\!\equiv\!&\Psi_{(-2,-1)}(z) 
\!=\!
\left( 
e^{{i\over 2}H} e^{-{1\over 2}\sigma}
-\frac{1}{\sqrt{2}}e^{-{i\over 2}H}\partial\xi e^{-{3\over 2}\sigma}
\right) e^{-{i\over 2}x-{1\over 2}\phi} \ ,
\eea
and the NS-sector operators 
\beq
u(z)\equiv\Psi_{(-1,-3)}(z)=x^2,~~~ 
v(z)\equiv\Psi_{(-3,-1)}(z)=y^2,~~~ 
w(z)\equiv\Psi_{(-2,-2)}(z)=xy \ ,
\eeq
given by 
\bea
u(z)
&=&
\left(-e^{-i H}e^{-\sigma}
+{i\over\sqrt{2}}c\partial\xi\partial(x-i\phi)e^{-2\sigma} 
-\sqrt{2}c(\partial^2\xi-\partial\xi\partial\sigma)e^{-2\sigma}
\right)e^{ix-\phi} \ ,\
\nn
w(z)
&=&\left(
{2\sqrt{2}\over 3}i\partial H\,e^{-\sigma}
+{i\over 3}c\partial\xi
\left[\partial(x+i\phi)e^{-iH}-\partial(x-i\phi)e^{iH}\right]
e^{-2\sigma}\right. \nn 
&&\left.+c(\partial^2\xi-\partial\xi\partial\sigma)(e^{iH}+e^{-iH})
e^{-2\sigma}
-c\partial\xi\partial(e^{iH}+e^{-iH})e^{-2\sigma}\phantom{\frac{2}{2}}\right)e^{-\phi} \ ,\
\nn
v(z)
&=&
\left(-e^{i H}e^{-\sigma}
+{i\over\sqrt{2}}c\partial\xi\partial(x+i\phi)e^{-2\sigma} 
+\sqrt{2}c(\partial^2\xi-\partial\xi\partial\sigma)e^{-2\sigma}
\right)e^{-ix-\phi} \ .
\end{eqnarray}
One has the multiplication rule
\begin{equation}
(\Psi_{(r,s)}\Psi_{(r^{\prime},s^{\prime})})(z)\sim
\Psi_{(r+r^{\prime}+1,s+s^{\prime}+1)}(z) \ ,
\end{equation}
where $\sim$ indicates that the right hand side could be multiplied by a vanishing constant.
The left sector ring of  spin zero, ghost number zero BRST invariant operators
is generated by the elements $x$ and $y$
\begin{eqnarray}\label{ringrel}
 \Psi_{(r,s)}=x^{-s-1}y^{-r-1},\quad\quad r,s\in{\mathbb Z}_- \ . 
\end{eqnarray}
Similarly, one can construct the ring in the right sector.
In order to construct the ground ring, we combine the left and right sectors
with the same left and right Liouville momenta.
We define
\begin{eqnarray}\label{ringelements}
  \begin{array}{ccc}
a_{ij}=\left(\begin{array}{cc}x\bar x&x\bar y\\
                         y\bar x&y\bar y \end{array}\right),
&\quad\quad&b_{ij}=\left(\begin{array}{ccc} u\bar u & u\bar w & u\bar v\\
                           w\bar u & w\bar w & w\bar v\\
                           v\bar u & v\bar w & v\bar v\end{array}\right) \ .
\end{array}
\end{eqnarray}
Next we need to impose the GSO projection.


\subsubsection{Type IIB}

We project with respect to $S^+(z)$ defined by
(\ref{Sp}) in the left moving sector and
with respect to 
$\bar{S}^+(\bar{z})$ defined by (\ref{bSp}) in the right moving sector.
The ground ring elements that survive the GSO projection are
\begin{eqnarray}
\label{0AR1}
  a_{ij}=\left(\begin{array}{cc}{\bf x\bar x}&\\& \end{array}\right),
\quad b_{kl}=\left(\begin{array}{ccc} u\bar u&&{\bf u\bar v}\\&&\\{\bf v\bar u}&&
{\bf v\bar v}
\end{array}\right) \ ,
\end{eqnarray}
where we denoted
by bold letters the generators of the ground ring. For instance, $b_{11}$ is generated as 
$a_{11}^2=b_{11}$.
In addition, we have to check locality with respect to the spectra we found in section
\ref{secIIb}.

\noindent
{\it ``Momentum Background, IIB$_m$''}

\noindent
In this case we find that all the elements surviving the GSO projection
are local with respect to the spectrum.
Thus, the ground ring is generated by four elements   $a_{11}, b_{33}, b_{13},b_{31}$
with the relation
 \begin{eqnarray}
\label{conz2}
  (a_{11})^2 b_{33}-b_{13}b_{31}=0 \ .
\eea
Note that 
$a_{11}$ is in the RR sector while the $b_{ij}$ are in the NSNS sector.
Note also that $a_{11}$ and $b_{33}$ 
are  momentum operators, while  
$b_{13},b_{31}$ are winding operators.

\noindent
{\it ``Winding Background, IIB$_w$''}

\noindent
Choosing the NSNS winding modes, however, we find that they are not local
with respect to the ground ring generator $a_{11}$. We conclude that in this
case the ground ring is generated by the elements 
$b_{11},b_{13},b_{31}$ and $b_{33}$, satisfying the 
relation
\begin{eqnarray}
\label{coni1}
  b_{11} b_{33}-b_{13}b_{31}=0 \ .
\eea

\subsubsection{Type IIA}

We project with respect to $S^+(z)$ in the left moving sector and
with respect to 
$\bar{S}^-(\bar{z})$ defined by (\ref{bSm}) in the right moving sector.
The ground ring elements that survive the GSO projection are
\begin{eqnarray}
\label{0AR2}
a_{ij}=\left(\begin{array}{cc} & {\bf x\bar y} \\&
              \end{array}\right),
\quad b_{kl}=\left(\begin{array}{ccc} {\bf u\bar u}&& u\bar v\\&&\\{\bf v\bar u}&&
{\bf v\bar v}
\end{array}\right) \ ,
\end{eqnarray}
where again we denoted
by bold letters the generators of the ground ring.

\noindent
{\it ``Momentum Background, IIA$_m$''}

\noindent
Checking locality with respect to the spectrum found in section \ref{secIIa}, we find that
$a_{12}$ fails to be local, whereas the remaining generators are local. 
In analogy to the type IIB case above, we find that the 
ground ring is generated by four elements  $b_{11}, b_{13}, b_{31},b_{33}$
with the relation
\begin{eqnarray}
\label{coni2}
  b_{11} b_{33}-b_{13}b_{31}=0 \ .
\eea

\noindent
{\it ``Winding Background, IIA$_w$''}

\noindent
Here, the ground ring is generated by four elements   $a_{12}, b_{31}, b_{11},b_{33}$
with the relation
 \begin{eqnarray}
\label{conz3}
  (a_{12})^2 b_{31}-b_{11}b_{33}=0 \ .
\eea

\noindent
{\it Ground Ring Singularity}

The (complex) equations (\ref{conz2}) and  (\ref{conz3})
define a singular Calabi-Yau 3-fold known as the
suspended pinch point singularity, whereas the (complex) equations
(\ref{coni1}) and (\ref{coni2}) define the ubiquitous conifold.

\subsubsection{Deformation of the Singularity}

\noindent
{\it Conifold Singularity}

Let us  first discuss the deformation of the conifold equation
(\ref{coni1}) and (\ref{coni2}).
A deformation consistent with the quantum numbers, i.e. the $x$ and $\phi$
momenta and the periodicity sector,  of $b_{11}b_{33}$ ($b_{13}b_{31}$)
is given by
\begin{equation}
\label{defconi1}
 b_{11} b_{33}-b_{13}b_{31}=\mu^2 \ ,
\end{equation}
where $\mu$ couples to the zero $x$-momentum RR operator in the respective theory.
In type IIB$_w$ it is
\beq
\label{RR1}
V_{RR}^{IIB} = 
e^{-\frac{\sigma}{2}-\frac{\bar{\sigma}}{2}+\frac{i}{2}\left(H+\bar{H}\right)
+\phi} \ ,
\eeq
while in type IIA$_m$ it reads 
\beq
\label{RR2}
V_{RR}^{IIA} = 
e^{-\frac{\sigma}{2}-\frac{\bar{\sigma}}{2}+\frac{i}{2}\left(H-\bar{H}\right)
+\phi} \ .
\eeq
We therefore suggest that the type IIB$_w$ and  IIA$_m$ theories deformed by
these RR operators are equivalent to the topological B-model on the
deformed conifold. 

Let us perform some consistency checks of the proposal.
Consider first the sphere partition function.  
In \cite{Kutasov:1990ua} the sphere partition function of the $2d$ type II string has been 
argued to vanish.
The reason is the existence of four fermionic zero modes, compared
to the two in the $N=1$ Liouville system where the sphere partition function is finite.
We expect a non-vanishing sphere partition
function of the corresponding topological B-model, which implies
a lift or projection of the two extra fermionic zero modes.
Indeed, the RR deformation we put breaks $N=2$ supersymmetry
on the worldsheet and therefore we expect the additional zero modes to be lifted.
Hence, we can expect a non-vanishing sphere partition function
also on the non-critical string side of our correspondence.

Next, let us compare the one-loop partition
functions of the deformed  type IIB$_w$ and  IIA$_m$ theories, with the 
one-loop partition of the topological B-model on the
deformed conifold.
The one-loop partition function of
 the B-model is given by $-\frac{1}{12}\log{\hat{\mu}}$,
where $\hat{\mu}$ is the deformation parameter of the conifold.
According to (\ref{defconi1}),  $\hat{\mu}= \mu^2$, and therefore
\beq
 Z_{B-model}= -\frac{1}{6}\log{\mu} \ .
\eeq
The calculation of the one-loop partition function in the type II non-critical string, 
on the other hand,
is a rather subtle issue. 
It was recently proposed 
\cite{Seiberg:2005bx} that the correct result is given by the regularized
sum over the physical spectrum, i.e. the spectrum obtained after imposing the 
Dirac constraint. 
However, we would like to propose that in the type IIA/B models at hand
the one-loop partition sum is obtained, as usually done, by summing over the spectrum
without imposing the Dirac constraint.

Indeed, if we apply this prescription for the spectra (\ref{iibw}) and
(\ref{iiam}) we find
\beq
Z_{IIA_m, IIB_w}=-\frac{1}{6}\log{\mu} \ ,
\eeq
with the $\log{\mu}$ giving the Liouville volume, as determined by the Liouville
momentum one operators (\ref{RR1}) and (\ref{RR2}).
With this prescription we find agreement between the one-loop partition functions
of the topological B-model and the deformed  type IIA/B theories.
Note, that if instead we summed only over the physical spectra (\ref{iibwd})
and (\ref{iiamd}),
we would get for the type  IIA/B theories $-\frac{1}{24}\log{\mu}$.
It remains an important open problem to calculate
the one-loop partition function in detail to verify the prescription.

\noindent
{\it Suspended Pinch Point Singularity}

\noindent
Consider next the singularities (\ref{conz2}) and  (\ref{conz3}), which are of the
type
\beq
\label{ssp}
x^2y -zw = 0 \ .
\eeq
The singularity of (\ref{ssp}) at the origin, can be resolved once by
changing coordinates to $z' = z/x$. $z'$ is a coordinate on a 2-sphere
that we grow at the origin.
With this change of variables equation  (\ref{ssp}) becomes
\beq
\label{conifold}
xy -z'w = 0 \ 
\eeq
which we recognize as the conifold equation.

We can deform the suspended pinch point singularity in two ways.
One is by
\beq
\label{sing}
x^2y-zw = \hat{\mu} \ ,
\eeq
and the second by 
\beq
\label{sing1}
x^2y-zw = \mu x \ .
\eeq
The latter deformation has been discussed
in \cite{Aganagic:2001ug}.
As we will show later, the deformation (\ref{sing1}) arises naturally
in the framework of the corresponding quiver gauge theories, where $\mu$ is related
to the dynamical scale $\Lambda$  and to the glueball superfield $S$.
It is also related to the deformation of the conifold equation
(\ref{conifold}) by the change of variables 
discussed above.
Locally the deformed space looks like
$T^*S^3$, where $\mu$ is the scale of $S^3$. Globally
it is different since, in particular,
$h^{1,1} =1$ for the deformed singularity (\ref{sing1}).

Consider the two types of deformation (\ref{sing}) and (\ref{sing1})
in the type IIB$_m$ and type IIA$_w$ theories.
Let us start with the deformation  (\ref{sing}).
In type IIB$_m$ the ground ring relation (\ref{conz2}) is deformed
as 
 \begin{eqnarray}
\label{conz44}
  (a_{11})^2 b_{33}-b_{13}b_{31}=\mu^2 \ ,
\eea
where $\mu$ couples to the RR operator
(\ref{RR1}).

Similarly,  in type IIA$_w$  the ground ring relation is deformed
as
 \begin{eqnarray}
\label{conz4}
  (a_{12})^2 b_{31}-b_{11}b_{33}=\mu^2 \ ,
\eea
where $\mu$ couples to the RR operator (\ref{RR2}).

We therefore suggest that the type IIB$_m$ and  IIA$_w$ theories deformed by
these RR operator are equivalent to the topological B-model on the
deformed suspended pinch point singularity (\ref{sing}). 

Again, we can compute the one-loop partition function by summing over the spectra
and get
\beq
Z_{IIA_w, IIB_m}=-\frac{1}{6}\log{\mu} \ .
\eeq
However, we do not know what is the one-loop partition function of the topological 
B-model on the deformed singularity 
(\ref{sing}). Following the proposed correspondence we suggest that it is given
by  $-\frac{1}{12}\log{\hat{\mu}}$, exactly as in the conifold.

Consider next the deformation  (\ref{sing1}).
In order to get in type IIB$_m$   a deformation of
the ground ring relation to 
 \begin{eqnarray}
\label{conz5}
  (a_{11})^2 b_{33}-b_{13}b_{31}=\mu a_{11}\ ,
\eea
we need that 
$\mu$ couples to the RR operator with $\frac{1}{2}$ $x$-momentum and
$\frac{3}{2}$ $\phi$-momentum. 
Looking at the spectra (\ref{iibd}) we see that there
is no such operator.
However, if we have the coupling 
$\hat{\mu}\int V_{NSNS}^{IIB} + \hat{\mu}\int V_{RR}^{IIB}$
where
\beq
V_{NSNS}^{IIB}= e^{-\sigma-\bar{\sigma}+\frac{i}{2}\left(
X_L+X_R\right)+\frac{\phi}{2}},~~~~~
V_{RR}^{IIB} = 
e^{-\frac{\sigma}{2}-\frac{\bar{\sigma}}{2}+\frac{i}{2}\left(H+\bar{H}\right)
+\phi} \ ,
\eeq
then the perturbation $\mu\int V_{NSNS}^{IIB}\int V_{RR}^{IIB}$
with $\mu=\hat{\mu}^2$ satisfies the requirements.

Similarly,  in type IIA$_w$  
in order to get a deformation of
the ground ring relation to
 \begin{eqnarray}
\label{conz6}
  (a_{12})^2 b_{31}-b_{11}b_{33}=\mu a_{12}\ ,
\eea
we suggest the
coupling $\hat{\mu}\int V_{NSNS}^{IIA}+\hat{\mu}\int V_{RR}^{IIA}$
such that $\mu=\hat{\mu}^2$,
and
\beq
V_{NSNS}^{IIA}= e^{-\sigma-\bar{\sigma}+\frac{i}{2}\left(
X_L-X_R\right)+\frac{\phi}{2}},~~~~~
V_{RR}^{IIA} = 
e^{-\frac{\sigma}{2}-\frac{\bar{\sigma}}{2}+\frac{i}{2}\left(H-\bar{H}\right)
+\phi} \ .
\eeq
We therefore propose  that the type IIB$_m$ and  IIA$_w$ theories deformed by the above
operators are equivalent to the topological B-model on the
deformed suspended pinch point singularity (\ref{sing1}). 

Again, we can compute the one-loop partition function by summing over the spectra,
and get
\beq
\label{result}
Z_{IIA_w, IIB_m}=-\frac{1}{12}\log{\mu} \ ,
\eeq
where $-\frac{1}{6}$ is the regularized summation over the spectra, and the Liouville
volume is $\log{\hat{\mu}}=\frac{1}{2}\log{\mu}$. To compare this result to the 
one-loop partition function of the B-model on the deformed suspended pinch
point (\ref{sing1}), we recall that in the patch $x \ne 0$ the geometry is described by
the conifold equation. Hence, we expect the usual conifold result $-\frac{1}{12}\log{\mu}$.
This will be justified more thoroughly in the last section where we also analyze
the corresponding quiver gauge theory.

Note, that we can perturb by a single RR operator, as well.
However, since the required $\phi$ momentum is bigger than $\frac{Q}{2}$,
this operator is in the ``non-local'' regime
 \cite{Seiberg:1990eb}.
In type IIB$_m$  such an operator is 
\beq
\label{RR3}
V_{RR~non-local}^{IIB}= 
e^{-\frac{\sigma}{2}-\frac{\bar{\sigma}}{2}-\frac{i}{2}\left(H+\bar{H}\right)
+\frac{i}{2}\left(X_L+X_R\right)+\frac{3}{2}\phi} \ ,
\eeq
while
in type IIA$_w$  
it reads
\beq
V_{RR~non-local}^{IIA}= 
e^{-\frac{\sigma}{2}-\frac{\bar{\sigma}}{2}-\frac{i}{2}\left(H-\bar{H}\right)
+\frac{i}{2}\left(X_L-X_R\right)+\frac{3}{2}\phi} \ .
\eeq
One may try to understand such a deformation,
with  the operator
inserted in the worldsheet path integral being 
$\int a_{11}^{-1}V_{RR~non-local}^{IIB} \sim e^{2\phi}+...$ in type IIB, and
similarly  $a_{12}^{-1}V_{RR~non-local}^{IIA}$ in type IIA.

It is amusing to check the 
one-loop partition functions.
Indeed, we will get the required result if we take the Liouville
volume to be  $\frac{1}{2}\log{\mu}$, which may be plausible
since the perturbations
$a_{11}^{-1}V_{RR~non-local}^{IIB}$ and $a_{12}^{-1}V_{RR~non-local}^{IIA}$ 
carry Liouville momentum two.

\subsection{$\cQ$ Projection}
\label{specsec}

The space-time supercharge  $\cQ$, obtained by the closed contour integral
of the supercurrent $S$ 
\beq
\cQ = \oint \frac{dz}{2 \pi i} S(z) \ ,
\eeq
is 
nilpotent $\cQ^2=0$.
Hence, $\cQ$ can be considered as a BRST like charge. 
One can consider a projected 
theory
obtained by restricting to ${\rm ker}\; \cQ$, with
physical operators satisfying
\beq
\label{Pr}
\cQ |phys\rangle=0 \ .
\eeq
The projected theories have been suggested as possible
topological theories \cite{Kutasov:1991pv,DiFrancesco:1991ud}.
Note that these theories are also candidates for a topological
 B-model correspondence, 
since the ground ring is invariant under the $\cQ$-projection \cite{Bouwknegt:1991am}.
In this section we will consider the spectra of these theories.
However, the results cast doubts on the consistency 
of these theories. Still, these results must be taken with a grain of salt.
More conclusive statements require a fully-fledged derivation of the
partition function for these theories.

We will first discuss the differences between the projections by 
$\cQ^+$ and $\cQ^-$ in the various chiral sectors. Then we will combine 
the $\cQ$-projected sectors to find the spectra of the projected
type IIA/B theories. 

In the NS sector the tachyon vertex operators $T_k$ are of the
form (\ref{Tk}).
Projecting onto the operators that satisfy $\cQ^+ T_k =0$, with
$\cQ^+$ given by the holomorphic integral of the current (\ref{Sp})
implies
\begin{equation}
\cQ^+ T_k=0 \to  \;\; k \in Z + \frac{1}{2} > 0 \ .
\end{equation}
On the other hand, using $\cQ^{-}$ for the projection gives the opposite condition
\begin{equation}
\cQ^- T_k=0 \to  \;\;  k \in Z + \frac{1}{2} < 0 \ .
\end{equation}
In the Ramond sector we have the vertex operators $V_k$ of the
form (\ref{Vk}),
with $\epsilon=\pm 1$.
The states of the $\epsilon=1$ sector, which are invariant under $\cQ^+$,
satisfy
\begin{equation}
\cQ^+ V_k^{\epsilon=1}=0 \to  \;\; k \in Z  \ge  0 \ ,
\end{equation}
whereas for the $\epsilon=-1$ sector we find
\begin{equation}
\cQ^+ V_k^{\epsilon=-1}=0 \to  \;\; k \in Z +\frac{1}{2}  >  0 \ .
\end{equation}
The operator $\cQ^-$ again acts in the opposite way and
the $\cQ^-$ invariant operators are given by
\begin{eqnarray}
\cQ^- V_k^{\epsilon=1}=0 & \to & k \in Z +\frac{1}{2}  <  0 \ , \\ \nonumber
\cQ^- V_k^{\epsilon=-1}=0 & \to & k \in Z \le  0 \ .
\end{eqnarray}

Let us now combine the left and right movers to get
the physical closed string states on the $\cQ$-projected theories. 
In the following we will present the results after imposing the
Dirac constraint. In the appendix we list the spectra before imposing 
it.
First we consider the theories obtained by using just the 
holomorphic $\cQ$ operator.
We obtain the following spectra consistent with the Dirac constraint
\begin{itemize}
\item{IIB$_m^{\cQ}$:}
\begin{eqnarray}
NSNS &  e^{-\sigma-\bar{\sigma}+i\left(n+\frac{1}{2}\right)\left(
X_L+X_R\right)} \ , & n \in Z \ge 0\ , \\ \nonumber
RR &
e^{-\frac{\sigma}{2}-\frac{\bar{\sigma}}{2}+\frac{i}{2}\left(H+\bar{H}\right)
+im\left(X_L+X_R\right)} \ , & m \in Z \ge 0 \ . 
\end{eqnarray}

\item{IIB$_w^{\cQ}$:}
\begin{eqnarray}
NSNS &  e^{-\sigma-\bar{\sigma}+i\left(n+\frac{1}{2}\right)\left(
X_L-X_R\right)} \ , & n \in Z \ge 0\ , \\ \nonumber
RR &
e^{-\frac{\sigma}{2}-\frac{\bar{\sigma}}{2}+\frac{i}{2}\left(H+\bar{H}\right)
+im\left(X_L+X_R\right)} \ , & m \in Z \ge 0 \ . 
\end{eqnarray}

\item{IIA$_m^{\cQ}$:}
\begin{eqnarray}
NSNS &  e^{-\sigma-\bar{\sigma}+i\left(n+\frac{1}{2}\right)\left(
X_L+X_R\right)} \ , & n \in Z \ge 0\ , \\ \nonumber
RR &
e^{-\frac{\sigma}{2}-\frac{\bar{\sigma}}{2}+\frac{i}{2}\left(H-\bar{H}\right)
+im\left(X_L-X_R\right)} \ , & m \in Z \ge 0 \ . 
\end{eqnarray}

\item{IIA$_w^{\cQ}$:}
\begin{eqnarray}
NSNS &  e^{-\sigma-\bar{\sigma}+i\left(n+\frac{1}{2}\right)\left(
X_L-X_R\right)} \ , & n \in Z \ge 0\ , \\ \nonumber
RR &
e^{-\frac{\sigma}{2}-\frac{\bar{\sigma}}{2}+\frac{i}{2}\left(H-\bar{H}\right)
+im\left(X_L-X_R\right)} \ , & m \in Z \ge 0 \ . 
\end{eqnarray}

\end{itemize}
We can also define theories by acting with both the
holomorphic $\cQ$ and the antiholomorphic $\bar{\cQ}$.
For these symmetric projections we find
\begin{itemize}
\item{IIB$_m^{\cQ,\bar{\cQ}}$:} 
\begin{eqnarray}
NSNS &  e^{-\sigma-\bar{\sigma}+i\left(n+\frac{1}{2}\right)\left(
X_L+X_R\right)} \ , & n \in Z \ge 0\ , \\ \nonumber
RR &
e^{-\frac{\sigma}{2}-\frac{\bar{\sigma}}{2}+\frac{i}{2}\left(H+\bar{H}\right)
+im\left(X_L+X_R\right)} \ , & m \in Z \ge 0 \ . 
\end{eqnarray}

\item{IIB$_w^{\cQ,\bar{\cQ}}$:} 
\begin{eqnarray}
RR &
e^{-\frac{\sigma}{2}-\frac{\bar{\sigma}}{2}+\frac{i}{2}\left(H+\bar{H}\right)
+im\left(X_L+X_R\right)} \ , & m \in Z \ge 0 \ . 
\end{eqnarray}

\item{IIA$_m^{\cQ,\bar{\cQ}}$:} 
\begin{eqnarray}
RR & e^{-\frac{\sigma}{2}-\frac{\bar{\sigma}}{2}+\frac{i}{2}\left(H-\bar{H}\right)
+im\left(X_L-X_R\right)} \ , & m \in Z \ge 0 \ .
\end{eqnarray}

\item{IIA$_w^{\cQ,\bar{\cQ}}$:} 
\begin{eqnarray}
NSNS &  e^{-\sigma-\bar{\sigma}+i\left(n+\frac{1}{2}\right)\left(
X_L-X_R\right)} \ , & n \in Z \ge 0\ , \\ \nonumber
RR &
e^{-\frac{\sigma}{2}-\frac{\bar{\sigma}}{2}+\frac{i}{2}\left(H-\bar{H}\right)
+im\left(X_L-X_R\right)} \ , & m \in Z \ge 0 \ . 
\end{eqnarray}

\end{itemize}

Note that none of the above theories features space-time fermions since
we are considering the $\cQ$ cohomology and the NSR states are $\cQ$-exact.

\noindent
{\it The Partition Function}

\noindent 
Let us turn now to the torus partition function.
Due to the absence of transverse oscillations we assume that the partition functions
in the $\cQ$-projected theories reduce
to sums over discrete momenta. For the theories to be consistent this sum 
has to be modular invariant.
However, as we show in the appendix, the sum fails to do so,
which suggests that the theories obtained
by the $\cQ$-projection may fail to be consistent.

\section{Quiver Gauge Theories}

We consider the quiver gauge theory obtained by D-branes wrappings in the suspended
pinch point geometry (see \cite{Morrison:1998cs,Uranga:1998vf,Franco:2005fd}).
Integrating out the D-brane open string degrees of freedom 
results in a deformed geometry interpreted as the backreaction
of the D-branes on the original geometry.
This  back reaction,
in the topological strings framework, affects the closed strings
by changing the periods of the holomorphic 3-form.
The Calabi-Yau space undergoes a transition, where holomorphic cycles
shrink and 3-cycles are opened up.
In the language of four-dimensional ${\cal N}=1$ supersymmetric gauge theory,
the D-branes wrapping the resolved geometry provide the UV description,
while the IR physics is described by the 
deformed geometry after the transition.

\subsection{Toric Geometry}

The suspended pinch point singularity is described by the
toric data

\begin{eqnarray}
&&\;\,\begin{array}{ccccc}\;z_1&z_2&\;z_3&\;\;z_4&\;z_5\end{array} \nonumber \\
  \left(\begin{array}{c}Q_1\\ Q_2\end{array}\right)&=&
\left(\begin{array}{rrrrr}\,1&\,0&\,1&\,-1&-1\\\,0&\,1&\,-1&\,1&-1\end{array}\right) \ .
\end{eqnarray}
These charges represent a $U(1)^2$  action on the coordinates $z_i \in
\mathbb C^5$.
The singular space is the symplectic quotient of  $\mathbb C^5$ by  $U(1)^2$ defined by the (D-term) equations
\begin{eqnarray}\label{ressp1}
 |z_1|^2+|z_3|^2-|z_4|^2- |z_5|^2&=&0 \ ,\nn
  |z_2|^2-|z_3|^2+|z_4|^2-|z_5|^2&=&0 \ ,
\end{eqnarray}
modulo the $U(1)^2$  action.
The resolved space is obtained by introducing two (FI) parameters  $t_1,t_2$ , such that
\begin{eqnarray}\label{ressp2}
  |z_1|^2+|z_3|^2-|z_4|^2- |z_5|^2&=&t_1 \ ,\nn
  |z_2|^2-|z_3|^2+|z_4|^2-|z_5|^2&=&t_2 \ .
\end{eqnarray}
In order to see that indeed the space we are considering is the 3-fold described by the
ground ring of the type II string, 
we will describe the toric singularity
in terms of a 
collection of polynomial equations, written in terms 
of $U(1)^2$-invariant monomials of the coordinates $z_i$. 
The monomials are generated by the four elements
\beq
\label{eq}
a_{22} = z_1z_2z_5,~~b_{11} = z_3z_4,~~b_{13}=z_1^2z_4z_5,~~b_{31}= z_2^2z_3z_5 ,
\eeq
and satisfy (\ref{conz2}).
In the geometric transition, an $S^3$ cycle opens up, and the relation (\ref{conz2})
is deformed to (\ref{sing1}).

\subsection{Quiver gauge theories}

The UV description of the quiver gauge theories is found by D-branes in the
resolved geometry.
Consider a system of D3-branes placed at the singularity (\ref{ressp1}), and
fractional D3-branes obtained by wrapping 
D5-branes on the two $\mathbb{P}^1$'s.
The quiver gauge theory reads

\beq
\label{quiv}
\begin{array}{cccc}
& SU(N_1) & SU(N_2) & SU(N_3)\\
X & \Box & & \bar\Box \\
{\tilde X} & \bar\Box & & \Box \\
Y & & \Box & \bar\Box   \\
{\tilde Y} & & \bar\Box & \Box   \\
Z & \Box & \bar\Box &  \\
{\tilde Z} & \bar\Box & \Box &  \\
\Phi & & &{\rm Adj.}  
\end{array}
\eeq

We will consider the setup where $N_1=N_3 \neq N_2$, obtained by placing 
$N_1=N_3$ D3-branes at the singularity and wrapping $N_2-N_1$
D5-branes on one $\mathbb{P}^1$.

The classical superpotential is
\begin{equation}
 W_{tree}=tr\left[\Phi(\tilde{Y}Y-\tilde{X}X)+ (Z\tilde{Z}X\tilde{X}-\tilde{Z}ZY\tilde{Y}) \right] \ .
\end{equation}

\subsubsection{Ring Structure}

The classical ring relations are obtained using the holomorphic
gauge invariants
\begin{equation}
 x=X\tilde{X}=Y\tilde{Y}, \quad y=Z\tilde{Z},\quad z=X\tilde{Y}\tilde{Z},\quad w=ZY\tilde{X}
\ ,
\end{equation}
which satisfy  $x^{2}y-zw=0$.
In order to see the quantum deformation
one considers $N_2=N_c$ and replaces the two other groups by $U(1)'s$
(D3-brane probes).
In terms of the meson
\begin{equation}
 M_{ij}=\left(\begin{array}[]{cc}Z\tilde{Z}&ZY\\
\tilde{Y}\tilde{Z} &	\tilde{Y}Y 
\end{array} \right) \ ,
 \end{equation}
the nonperturbative superpotential is given by
\begin{eqnarray}
W_{np} =(N_{c}-2)\left(\frac{\Lambda^{3N_c-2}}{det\,M} \right)^{\frac{1}{N_{c}-2}}+
tr\left[\Phi(M_{22}-\tilde{X}X)+(M_{11}X\tilde{X}-M_{12}M_{21}) \right] \ ,
\end{eqnarray}
which deforms the ring relation to
\begin{equation}
	x^{2}y-zw=\mu\, x,\quad 
\mu=\left(\Lambda^{3N_{c}-2}\right)^{\frac{1}{N_{c}-1}}.
 \end{equation}

\subsubsection{Holomorphic F-terms and B-model Free Energy}

Let us consider now the IR dynamics of the 
quiver gauge theory. 
After running the duality cascade for the quiver gauge theory, 
one ends up in the IR with 
a confining pure SYM gauge theory. 
In the suspended pinch point deformed geometry,  
we identify the size of $S^3$ with the
glueball superfield $S$.
The holomorphic F-terms of this theory ${\cal F}_{SYM}(S)$ as a function of  $S$ 
is 
\beq
{\cal F}_{SYM}(S) = F_{c=1}(R_{self-dual}) \ ,
\eeq
where $S\sim \mu$,
and
\beq
\label{part}
 F_{c=1}(R_{self-dual}) ={1\over 2}\mu^2 \log\mu -{1\over 12} \log\mu + 
{1\over 240}\mu^{-2}+\sum_{g>2} a_g \mu^{2-2g} \ ,
\eeq
where $a_g = \frac{B_{2g}}{2g(2g-2)}$.

We therefore propose that this is also the perturbative free energy of the topological  
B-model on the deformed suspended pinch point geometry (\ref{sing1}).
This seems very plausible, since locally the deformed suspended pinch point
singularity looks like
the deformed conifold $T^*S^3$, and the   perturbative free energy of the topological 
B-model on the deformed conifold is given by  (\ref{part}).
However, since the deformed conifold  and the deformed suspended pinch point geometries
are globally different, one should expect that the  free energy of the topological  
B-models on these geometries
will differ
non-perturbatively,

Returning now to the  
perturbed type II strings, as discussed in section (2.3.3),
we propose that in all cases described there, the perturbative free
energy is also given by  (\ref{part}), with the appropriate map of the
deformation parameters as discussed there.
We argued for the consistency of this proposal at tree level and one-loop.
Going beyond that will probably require a matrix model
description of the system, which is currently lacking\footnote{We refer the reader
to \cite{McGreevy:2003dn,Verlinde:2004gt,Takayanagi:2004ge} and the references therein 
for related results in this direction.}.
It would also be interesting to construct the target space geometry 
of the perturbed two-dimensional strings.

\section*{Acknowledgements}

We would like to thank D. Kutasov, N. Seiberg, A. Uranga and J. Walcher for valuable discussions. 
Y.O. would like to thank the theory division in CERN for hospitality
during the final stage of the work.
The work of H.I. was supported by the EPSRC grant GR/S73792/01.
\newline

\newpage

\appendix

\section{$\cQ$ projection}

In this appendix we list the spectra of the type II theories, which are
consistent with the holomorphic $\cQ$ projection. We do not impose the Dirac condition
here.

\begin{itemize}

\item{IIB$_m^{\cQ}:$}
\begin{eqnarray}
NSNS &  e^{-\sigma-\bar{\sigma}+i\left(n+\frac{1}{2}\right)\left(
X_L+X_R\right)} \ , & n \in Z \ge 0\ , \\ \nonumber
RR &
e^{-\frac{\sigma}{2}-\frac{\bar{\sigma}}{2}+\frac{i}{2}\left(H+\bar{H}\right)
+im\left(X_L\pm X_R\right)} \ , & m \in Z \ge 0 \ , \\ \nonumber 
 & e^{-\frac{\sigma}{2}-\frac{\bar{\sigma}}{2}-\frac{i}{2}\left(H+\bar{H}\right)
+i\left(n+\frac{1}{2}\right) \left(X_L+ X_R\right)} \ , & n \in Z \ge 0 \ ,
\end{eqnarray}

\item{IIB$_w^{\cQ}:$}
\begin{eqnarray}
NSNS &  e^{-\sigma-\bar{\sigma}+i\left(n+\frac{1}{2}\right)\left(
X_L-X_R\right)} \ , & n \in Z \ge 0\ , \\ \nonumber
RR &
e^{-\frac{\sigma}{2}-\frac{\bar{\sigma}}{2}+\frac{i}{2}\left(H+\bar{H}\right)
+im\left(X_L \pm X_R\right)} \ , & m \in Z \ge 0 \ , \\ \nonumber
 & e^{-\frac{\sigma}{2}-\frac{\bar{\sigma}}{2}-\frac{i}{2}\left(H+\bar{H}\right)
+i\left(n+\frac{1}{2}\right) \left(X_L- X_R\right)} \ , & n \in Z \ge 0 \ ,
\end{eqnarray}

\item{IIA$_m^{\cQ}:$}
\begin{eqnarray}
NSNS &  e^{-\sigma-\bar{\sigma}+i\left(n+\frac{1}{2}\right)\left(
X_L+X_R\right)} \ , & n \in Z \ge 0\ , \\ \nonumber
RR &
e^{-\frac{\sigma}{2}-\frac{\bar{\sigma}}{2}+\frac{i}{2}\left(H-\bar{H}\right)
+im\left(X_L \pm X_R\right)} \ , & m \in Z \ge 0 \ , \\ \nonumber
 & e^{-\frac{\sigma}{2}-\frac{\bar{\sigma}}{2}-\frac{i}{2}\left(H-\bar{H}\right)
+i\left(n+\frac{1}{2}\right) \left(X_L+ X_R\right)} \ , & n \in Z \ge 0 \ ,
\end{eqnarray}

\item{IIA$_w^{\cQ}:$}
\begin{eqnarray}
NSNS &  e^{-\sigma-\bar{\sigma}+i\left(n+\frac{1}{2}\right)\left(
X_L-X_R\right)} \ , & n \in Z \ge 0\ , \\ \nonumber
RR &
e^{-\frac{\sigma}{2}-\frac{\bar{\sigma}}{2}+\frac{i}{2}\left(H-\bar{H}\right)
+im\left(X_L \pm X_R\right)} \ , & m \in Z \ge 0 \ , \\ \nonumber
& e^{-\frac{\sigma}{2}-\frac{\bar{\sigma}}{2}-\frac{i}{2}\left(H-\bar{H}\right)
+i\left(n+\frac{1}{2}\right) \left(X_L- X_R\right)} \ , & n \in Z \ge 0 \ . 
\end{eqnarray}

\end{itemize}

\section{$\cQ,\bar{\cQ}$ projection}

In this appendix we have compiled the list of spectra consistent with
the combined holomorphic/antiholomorphic $\cQ,\bar{\cQ}$-projection.
Again, the Dirac condition is not imposed.

\begin{itemize}

\item{IIB$_m^{\cQ,\bar{\cQ}}:$}
\begin{eqnarray}
NSNS &  e^{-\sigma-\bar{\sigma}+i\left(n+\frac{1}{2}\right)\left(
X_L+X_R\right)} \ , & n \in Z \ge 0\ , \\ \nonumber
RR &
e^{-\frac{\sigma}{2}-\frac{\bar{\sigma}}{2}+\frac{i}{2}\left(H+\bar{H}\right)
+im\left(X_L+ X_R\right)} \ , & m \in Z \ge 0 \ , \\ \nonumber
 & e^{-\frac{\sigma}{2}-\frac{\bar{\sigma}}{2}-\frac{i}{2}\left(H+\bar{H}\right)
+i\left(n+\frac{1}{2}\right) \left(X_L+ X_R\right)} \ , & n \in Z \ge 0 \ ,
\end{eqnarray}

\item{IIB$_w^{\cQ,\bar{\cQ}}:$}
\begin{eqnarray}
RR &
e^{-\frac{\sigma}{2}-\frac{\bar{\sigma}}{2}+\frac{i}{2}\left(H+\bar{H}\right)
+im\left(X_L + X_R\right)} \ , & m \in Z \ge 0 \ ,
\end{eqnarray}

\item{IIA$_m^{\cQ,\bar{\cQ}}:$}
\begin{eqnarray}
RR &
e^{-\frac{\sigma}{2}-\frac{\bar{\sigma}}{2}+\frac{i}{2}\left(H-\bar{H}\right)
+im\left(X_L - X_R\right)} \ , & m \in Z \ge 0 \ ,
\end{eqnarray}

\item{IIA$_w^{\cQ,\bar{\cQ}}:$}
\begin{eqnarray}
NSNS &  e^{-\sigma-\bar{\sigma}+i\left(n+\frac{1}{2}\right)\left(
X_L-X_R\right)} \ , & n \in Z \ge 0\ , \\ \nonumber
RR &
e^{-\frac{\sigma}{2}-\frac{\bar{\sigma}}{2}+\frac{i}{2}\left(H-\bar{H}\right)
+im\left(X_L - X_R\right)} \ , & m \in Z \ge 0 \ , \\ \nonumber
& e^{-\frac{\sigma}{2}-\frac{\bar{\sigma}}{2}-\frac{i}{2}\left(H-\bar{H}\right)
+i\left(n+\frac{1}{2}\right) \left(X_L- X_R\right)} \ , & n \in Z \ge 0 \ . 
\end{eqnarray}

\end{itemize}

\section{$\cQ$-Projection and Modular Properties}

In the following we will only discuss the momentum-type theories but similar arguments
apply to the winding-type theories.
Consider first the type IIB theory with the $\cQ$-projection acting
only on the left-moving side. 
The total partition sum splits into sums over the 
possible spin structures. Explicitly, these sums are given by
\begin{eqnarray}
Z_{NSNS} & = & \sum_{s \ge 0, m \in Z} q^{\frac{1}{2}\left(s+\frac{1}{2}\right)^2}
\bar{q}^{\frac{1}{2}\left(s-2m+\frac{1}{2}\right)^2} \ , \nonumber \\
Z_{RR}^{\epsilon_L=\epsilon_R=1} & = & \sum_{s \ge 0, m \in Z} q^{\frac{1}{2}s^2}
\bar{q}^{\frac{1}{2}\left(s-2m\right)^2} \ , \nonumber \\
Z_{RR}^{\epsilon_L=\epsilon_R=-1} & = & \sum_{s \ge 0, m \in Z} q^{\frac{1}{2}\left(s+\frac{1}{2}\right)^2}
\bar{q}^{\frac{1}{2}\left(s-2m+\frac{1}{2}\right)^2} \ , \nonumber \\
\end{eqnarray}
Since the spectrum is given by the $\cQ$ cohomology, the space-time fermions are projected out.
As usual, they are $\cQ$-exact states.

To exhibit the properties of the above sums under modular transformations we will rewrite them in terms
of theta functions $\theta_i(\tau)$. This leads to the following expressions
\begin{eqnarray}
Z_{NSNS}=Z_{RR}^{\epsilon_L=\epsilon_R=-1}=
& = & \frac{1}{4}\left|\theta_2\right|^2 \ , \nonumber \\
Z_{RR}^{\epsilon_L=\epsilon_R=1} & = &\frac{1}{4}\left( \bar{\theta}_3+\bar{\theta}_4\right)
+\frac{1}{4}\left(\left|\theta_3\right|^2+\left|\theta_4\right|^2\right) \ .
\end{eqnarray}
From these expressions it is obvious that the sum over the spin structures is not modular invariant.
The most important problem is the sum over the $RR$ sector with $\epsilon_L=\epsilon_R=1$ which includes
a zero momentum mode. The sum of this mode over the right sector gives rise to the problematic
$ \bar{\theta}_3+\bar{\theta}_4$ contribution. 

\noindent
{\it $\cQ$-projection on both sides}

\noindent
If we follow the discussion of section \ref{specsec} and apply the $\cQ$-projection
on both sides in the type IIB theory, we get the following sums
\begin{eqnarray}
Z_{NSNS} & = & \sum_{m,n \ge 0} \left[ q^{\frac{1}{2} \left(2m+1+\frac{1}{2}\right)^2}
\bar{q}^{\frac{1}{2}\left(2n+1+\frac{1}{2}\right)^2}+
q^{\frac{1}{2} \left(2m+\frac{1}{2}\right)^2} 
\bar{q}^{\frac{1}{2}\left(2n+\frac{1}{2}\right)^2} \right]  \nonumber \\
& = & \frac{1}{4} \left|\theta_2\right|^2 -
\sum_{m,n \ge 0}\left[  q^{\frac{1}{2} \left(2m+1+\frac{1}{2}\right)^2}
\bar{q}^{\frac{1}{2}\left(2n+\frac{1}{2}\right)^2}+
q^{\frac{1}{2} \left(2m+\frac{1}{2}\right)^2} 
\bar{q}^{\frac{1}{2}\left(2n+1+\frac{1}{2}\right)^2} \right]  \nonumber \\
& = & Z_{RR}^{\epsilon_L=\epsilon_R=-1} \ , \nonumber \\
Z_{RR}^{\epsilon_L=\epsilon_R=1} & = &  \sum_{m,n \ge 0} \left[ q^{\frac{1}{2} \left(2m\right)^2}
\bar{q}^{\frac{1}{2}\left(2n\right)^2}+
q^{\frac{1}{2} \left(2m+1\right)^2} 
\bar{q}^{\frac{1}{2}\left(2n+1\right)^2} \right]  \nonumber \\
& = & \frac{1}{4} + \frac{1}{8}\left[\theta_3+\bar{\theta}_3+
\theta_4+\bar{\theta}_4+\left|\theta_3\right|^2+\left|\theta_4\right|^2 \right] \ , \nonumber \\
\end{eqnarray}
We can argue that even without expressing the remaining sums in terms
of theta-functions, the sum over the spin structures fails to be modular
invariant. This is because, in the above expression, the purely left (right) moving sums in the RR sector are not modular 
covariant. Thus, for the total sum to be modular covariant compensating purely left (right) 
moving sums are required. However, since the remaining sums are over half-integer
momenta there are no such purely left (right) moving sums. Therefore, the total sum is 
not modular covariant.

A similar analysis shows that the partition functions of
the $\cQ$-projected type IIA theories are not modular invariant.
Indeed, we can write the sum of the $\cQ$-projected type IIA/B partition functions in a fairly
nice form as
\begin{eqnarray}
Z_{IIA}+Z_{IIB} & = & \frac{1}{2} + \frac{1}{2} \left|\theta_2\right|^2
\frac{1}{4}\left(\left|\theta_3\right|^2+\left|\theta_4\right|^2\right)
+\frac{1}{4}\left( \theta_3 + \bar{\theta}_3 +\theta_4 + \bar{\theta}_4 \right) \ .
\end{eqnarray}
Note that this expression is not modular covariant. Since we expect either both or neither of them
to be modular covariant, the above expression provides another hint that
these theories may lack a modular invariant spectrum and are potentially inconsistent.

\end{document}